\documentclass[MSNbibl,number,citesort,dvips]{arxbj}
\usepackage{graphicx}
\aid{0}
\volume{20}
\issue{2}
\pubyear{2014}
\firstpage{676}
\lastpage{696}
\doi{10.3150/12-BEJ502} %kopijuoti is PTS

\makeatletter

\newcommand{\pa}{\mathrm{pa}}
\newcommand{\nei}{\operatorname{ne}}
\newcommand{\an}{\operatorname{an}}
\newcommand{\ant}{\operatorname{ant}}

\newcommand{\cip}{\perp\!\!\!\perp}
\newcommand{\cd}{\mid}
\newcommand{\nn}[0]{\hspace*{.7em}}

\newcommand{\ful}{\, \frac{ \nn\nn\;}{ \nn\nn
}\,}

\newcommand{\fla}{\hspace{.05em} \prec
\!\!\!\!\!\frac{\nn\nn}{\nn}\hspace{.20em}}

\newcommand{\fra}{\hspace{.15em} \frac{\nn
\nn}{\nn
}\!\!\!\!\! \succ\! \hspace{.35ex}}

\newcommand{\arc}{\hspace{.06em} \prec
\!\!\!\!\!\frac{\nn\nn}{\nn}\!\!\!\!\!
\succ\! \hspace{.35ex}}

\newtheorem{prop}{Proposition}
\newtheorem{coro}{Corollary}
\newtheorem{lemma}{Lemma}
\newtheorem{theorem}{Theorem}

\makeatother

\begin{document}
\begin{frontmatter}

\title{Markov properties for mixed graphs}
\runtitle{Markov properties for mixed graphs}

\begin{aug}
%%%% inicialai - be tarpu
\author[1]{\fnms{Kayvan} \snm{Sadeghi}\thanksref{1}\ead[label=e1]{kayvans@andrew.cmu.edu}} \and
\author[2]{\fnms{Steffen} \snm{Lauritzen}\corref{}\thanksref{2}\ead[label=e2]{steffen@stats.ox.ac.uk}}
\runauthor{K. Sadeghi and S. Lauritzen} %% auto
\address[1]{Department of Statistics, Baker Hall, Carnegie Mellon
University, Pittsburgh, PA 15213, USA.
\\\printead{e1}}
\address[2]{Department of Statistics, University of Oxford, 1 South
Parks Road, Oxford, OX1 3TG, United Kingdom. \printead{e2}}
\end{aug}

% HISTORY:
\received{\smonth{5} \syear{2012}}
\revised{\smonth{10} \syear{2012}}

% ABSTRACT
%
\begin{abstract}
In this paper, we unify the Markov theory of a variety of different
types of graphs used in
graphical Markov models by introducing the class of loopless mixed
graphs, and show that all
independence models induced by $m$-separation on such graphs are
compositional graphoids. We
focus in particular on the subclass of ribbonless graphs which as
special cases include undirected
graphs, bidirected graphs, and directed acyclic graphs, as well as
ancestral graphs and summary graphs.
We define maximality of such graphs as well as a pairwise and a global
Markov property. We prove that
the global and pairwise Markov properties of a maximal ribbonless graph
are equivalent for any
independence model that is a compositional graphoid.
\end{abstract}

% KEYWORDS
% visi is mazosios raides ir pagal abecele
%

\end{frontmatter}

%
%s1 #&#
\section{Introduction}
\label{secintro}
%s1.1 #&#
\subsection{Introduction and motivation} Graphical Markov models have
become widely used in recent years. The models use graphs to represent
conditional independence
relations for systems of random variables, with nodes of the graph
corresponding to random variables and edges representing dependencies.
Several classes of graphs with various independence interpretations
have been described in the literature. These range from undirected
graphs with simple separation for derivation of independencies \cite
{lau96} to various forms of mixed graphs \cite{kos02,ric02,wer11},
including chain graphs with several different separation criteria
\cite{fry90,cox93,koster97,and01,drt09}.

In spite of the differences among these graphs, their structural
similarities motivate an attempt to unify them. For this purpose, we
introduce the class of loopless mixed graphs and let them entail
independence models using the same separation criterion,
$m$-separation. This unification covers many graphical independence
models in the literature
with some independence models for chain graphs forming a notable
exception; see Section~\ref{secspecial} for further details. We show
that any independence model generated by $m$-separation in a loopless
mixed graph is a compositional graphoid. This ensures that certain
intuitive methods of reasoning are indeed valid for such graphs, as
they in some sense behave as ordinary undirected graphs.

A common motivation for defining MC-graphs \cite{kos02}, summary
graphs \cite{wer11}, and ancestral graphs \cite{ric02}, is to
represent independence relations implied by marginalisation over and
conditioning on sets of variables satisfying the Markov property of a
directed acyclic graph (DAG). The focus of our study is on a subclass
of loopless mixed graphs which we shall term \emph{ribbonless}. The
class of ribbonless graphs is sufficiently rich to serve the same
purpose: these graphs are obtained by a simple modification of MC
graphs derived from a DAG after marginalisation and conditioning; and
it contains summary graphs and ancestral graphs as special cases.

For ribbonless graphs, we define global and pairwise Markov properties,
the latter being associated with interpreting missing edges in the
graph as representing conditional independencies. We prove as our main
result that a
compositional graphoid independence model over a maximal ribbonless
graph satisfies the global Markov property if and only if it satisfies
the pairwise Markov property. This ensures that the independence models
represented by such graphs are generated by their missing edges, which
again supports the direct visual intuition.
%s1.2 #&#
\subsection{Some early results on Markov properties}
The concepts of
pairwise and global Markov properties for undirected graphs were
introduced in \cite{ham71} in the context of random fields and shown
to be equivalent for positive densities. Alternative proofs were later
given independently by several authors, for example \cite
{gri73,bes74}; see also \cite{clifford90}. An abstract variant of
this theorem was proven in \cite{pea88} for independence models
satisfying graphoid axioms as these are satisfied by probabilistic
distributions with positive densities; see also \cite{stu89} and
\cite{gei90}. Independence models for undirected graphs were
discussed comprehensively in Chapter~3 of \cite{lau96}.

A global Markov property that uses the $m$-separation criterion and a
pairwise Markov property were defined in \cite{ric02} for maximal
ancestral graphs without considering conditions under which they are
equivalent. We use a generalisation of these Markov properties for
maximal ribbonless graphs, which contains maximal ancestral graphs as a
subclass, and prove their equivalence for compositional graphoids. This
has been mentioned as a conjecture in \cite{kan09}.

%s1.3 #&#
\subsection{Structure of the paper} In the next section, we
introduce the basic concepts of graph theory, general and probabilistic
independence models, and compositional graphoids.

In Section~\ref{seclmg}, we introduce the class of loopless mixed
graphs and additional graph theoretical definitions special to mixed
graphs. We also associate the $m$-separation criterion to this class,
and prove for any loopless mixed graph that the independence model
induced by $m$-separation is a compositional graphoid.

In Section~\ref{secspecial}, we introduce the class of ribbonless
graphs and the concept of anterior graphs. We describe the relations
between these as well as subclasses of loopless mixed graphs that have
been discussed in the literature.

In Section~\ref{secmaximal}, we introduce the concept of maximality
by demanding that any additional edge will change the independence
model. It is shown that ribbonless graphs are not necessarily maximal,
and conditions for maximality are given.

In Section~\ref{secmarkov}, we define a pairwise and a global Markov
property for independence models for ribbonless graphs, and prove our
main result: that pairwise and global Markov properties are equivalent
for compositional graphoid independence models over maximal ribbonless graphs.

%In Sec. 7 we define probabilistic independence models and provide a
%short discussion on the results implied for probability distributions.
%s2 #&#
\section{Basic definitions and concepts}
\label{secbasic}
In this section, we introduce basic definitions and notation for
independence models, graphs, and compositional graphoids.
%s2.1 #&#
\subsection{Basic graph theoretical definitions}
A \emph{graph} $G$ is a triple consisting of a \emph{node} set or
\emph{vertex} set $V$, an \emph{edge} set $E$, and a relation that with
each edge associates two nodes (not necessarily distinct), called
its \emph{endpoints}. When nodes $i$ and $j$ are the endpoints of an
edge, they are
\emph{adjacent} and we write $i\sim j$. We say the edge is \emph
{between} its two
endpoints. We usually refer to a graph as an ordered
pair $G=(V,E)$. Graphs $G_1=(V_1,E_1)$ and $G_2=(V_2,E_2)$ are called
\emph{equal} if $(V_1,E_1)=(V_2,E_2)$. In this case we write $G_1=G_2$.

Notice that our graphs are \emph{labeled}, that is, every node is
considered as a different object. Hence, for example, graph $i\ful
j\ful k$ is not equal to $j\ful i\ful k$.

A \emph{loop} is an edge with
the same endpoints. \emph{Multiple edges} are edges with the
same pair of endpoints. A \emph{simple graph} has neither
loops nor multiple edges.

%The first node of the ordered pair is the \emph{tail} and the second
%is the \emph{head}.
A \emph{subgraph} of a graph $G_1$ is a graph $G_2$ such that
$V(G_2)\subseteq V(G_1)$ and $E(G_2)\subseteq E(G_1)$ and the
assignment of endpoints to edges in $G_2$
is the same as in $G_1$. An \emph{induced subgraph} by nodes
$A\subseteq V$ is a subgraph that contains all and only nodes in $A$
and all edges between two nodes in $A$. A subgraph induced by edges
$B\subseteq E$ is a subgraph that contains all and only edges in $B$
and all nodes that are endpoints of edges in $B$.

A \emph{walk} is a list $\langle v_0,e_1,v_1,\dots,e_k,v_k\rangle$
of nodes and edges such that for $1\leq i\leq k$, the edge $e_i$ has
endpoints $v_{i-1}$ and $v_i$. A \emph{path} is a walk with no
repeated node or edge.
If the graph is simple then the path can be uniquely determined by an
ordered sequence of node sets. Throughout this paper, we use node
sequences to describe paths even in graphs with multiple edges, as it
usually is apparent from the context which of multiple edges belong to
the path.
We say a path is \emph{between} the first and the last nodes of the
list in $G$. We
call the first and the last nodes \emph{endpoints} of the path and all
other nodes \emph{inner nodes}.

If $\pi_1=\langle i=i_0,i_1,\dots,i_n,h\rangle$ and $\pi_2=\langle
h,j_m,j_{m-1},\dots,j_0=j\rangle$ are paths, their \emph
{combination} $\pi_{12}=\pi_1\circ\pi_2$ is the path $\pi_{12}=
\langle i,\ldots,i_{p-1}, k,j_{q-1},\dots,j\rangle$, where
$k=i_p=j_q$ is the first node of $\pi_1$ which is on both paths. If
$k=h$ then $\pi_{12}$ is simply the concatenation of the two paths. In
general, the concatenation of two paths will be a walk and not a path
as the paths may intersect in more than one point.
%A graph is \emph{connected} if there is a path between any pair of
%nodes.
%
%In general, nodes connected by paths define a \emph{connected
%component} in the graph. Thus, a connected graph has only one
%connected component. In this paper we mostly focus on connected
%graphs, and by ``graph" we mean ``connected graph", unless otherwise
%stated.

A \emph{subpath} of a path $\pi$ is a path that can be considered a
subgraph of $\pi$ with the ordering associated with $\pi$.
%a path that can be considered as a subgraph of $\pi$ with the ordering
%associated with $\pi$.
A \emph{cycle} in a graph $G$ is a simple subgraph whose nodes can be
placed around a circle so that two nodes are
adjacent if they appear consecutively along the circle.

%s2.2 #&#
\subsection{Independence models}
An \emph{independence model} $\mathcal{J}$ over a set $V$ is a set of
triples $\langle X,Y\cd Z\rangle$ (called \emph{independence
statements}), where $X$, $Y$, and $Z$ are disjoint subsets of $V$ and $Z$
can be empty, and \mbox{$\langle\varnothing,Y\cd Z\rangle$} and $\langle
X,\varnothing\cd Z\rangle$ always being included in $\mathcal{J}$.
The independence statement \mbox{$\langle X,Y\cd Z\rangle$} is interpreted as
``$X$ is independent of $Y$ given $Z$''.

An independence model $\mathcal{J}$ over a set $V$ is a \emph
{semi-graphoid} if for disjoint subsets $A$, $B$, $C$, and $D$ of $V$,
it satisfies the four following properties:
\begin{enumerate}[3.]
\item$\langle A,B\cd C\rangle\in\mathcal{J}$ if and only if
$\langle B,A\cd C\rangle\in\mathcal{J}$ (\emph{symmetry});
\item if $\langle A,B\cup D\cd C\rangle\in\mathcal{J}$ then $\langle
A,B\cd C\rangle\in\mathcal{J}$ and $\langle A,D\cd C\rangle\in
\mathcal{J}$ (\emph{decomposition});
\item if $\langle A,B\cup D\cd C\rangle\in\mathcal{J}$ then $\langle
A,B\cd C\cup D\rangle\in\mathcal{J}$ and $\langle A,D\cd C\cup
B\rangle\in\mathcal{J}$ (\emph{weak union});
\item$\langle A,B\cd C\cup D\rangle\in\mathcal{J}$ and $\langle
A,D\cd C\rangle\in\mathcal{J}$
if and only if $\langle A,B\cup D\cd C\rangle\in\mathcal{J}$ (\emph
{contraction}).
\end{enumerate}
A semi-graphoid for which the reverse implication of the weak union
property holds is said to be a \emph{graphoid}, that is
\begin{enumerate}[5.]
\item[5.] if $\langle A,B\cd C\cup D\rangle\in\mathcal{J}$ and
$\langle A,D\cd C\cup B\rangle\in\mathcal{J}$ then $\langle A,B\cup
D\cd C\rangle\in\mathcal{J}$ (\emph{intersection}).
\end{enumerate}
Furthermore, a graphoid or semi-graphoid for which the reverse
implication of the decomposition property holds is said to be \emph
{compositional}, that is
\begin{enumerate}[6.]
\item[6.] if $\langle A,B\cd C\rangle\in\mathcal{J}$ and $\langle
A,D\cd C\rangle\in\mathcal{J}$ then $\langle A,B\cup D\cd C\rangle
\in\mathcal{J}$ (\emph{composition}).
\end{enumerate}
Notice that simple separation in an undirected graph will trivially
satisfy all of these properties, and hence compositional graphoids are
direct generalisations of independence models given by separation in
undirected graphs.

%s2.3 #&#
\subsection{Probabilistic conditional independence models}
The most
common independence models are induced by probability distributions.
Consider a set $V$ and a collection of random variables
$(X_\alpha)_{\alpha\in V}$ with state spaces $\mathcal{X}_\alpha,
\alpha\in V$ and joint distribution $P$. We let $X_A=(X_v)_{v\in A}$
etc. for each subset $A$ of $V$. For disjoint subsets $A$, $B$, and $C$
of $V$
we use the short notation $A\cip B\cd C$ to denote that $X_A$ is \emph
{conditionally independent of $X_B$ given $X_C$} \cite{daw79,lau96},
that is, that for any measurable $\Omega\subseteq\mathcal{X}_A$ and
$P$-almost all $x_B$ and $x_C$,
\[
P(X_A \in\Omega\cd X_B=x_B,
X_C=x_C)=P(X_A \in\Omega\cd
X_C=x_C).
\]
We can now induce an independence model $\mathcal{J}(P)$ by letting
\begin{displaymath}
\langle A,B\cd C\rangle\in\mathcal{J}(P) \quad \mbox{if and only if}\quad A\cip
B\cd C\qquad \mbox{w.r.t. }P.
\end{displaymath}
We say that an independence model $\mathcal{J}$ is \emph
{probabilistic} if there is a distribution $P$ such that $\mathcal{J}=
\mathcal{J}(P)$. We then also say that $P$ is \emph{faithful} to
$\mathcal{J}$.

Probabilistic independence models are always semi-graphoids \cite
{pea88}, whereas the converse is not necessarily true; see \cite
{stu89}. If $P$ has strictly positive density, the induced independence
model is also a graphoid; see, for example, Proposition~3.1 in \cite
{lau96}. If the distribution $P$ is a regular multivariate Gaussian
distribution, $\mathcal{J}(P)$ is a compositional graphoid. This
follows from the fact that for such a distribution
\[
A\cip B\cd C \quad \iff\quad k^{\alpha\beta}_{A\cup B\cup C}=0 \qquad \mbox{for
all } \alpha\in A,\beta\in B,
\]
where $k^{\alpha\beta}_{A\cup B\cup C}$ is the $\alpha\beta$ entry
in the concentration matrix of the distribution of $X_{A\cup B\cup C}$
and hence setwise conditional independence is directly determined by
nodewise conditional independence.

Probabilistic independence models with positive densities are not in
general compositional graphoids; this only holds for special types of
multivariate distributions such as the Gaussian mentioned above and,
say, the symmetric binary distributions used in \cite{wer09}.\vspace*{-2pt}

%s3 #&#
\section{Independence models for mixed graphs}\vspace*{-2pt}
\label{seclmg}
%In this section we discuss loopless mixed graphs and their induced
%independence models.
%s3.1 #&#
\subsection{Mixed graphs}\vspace*{-2pt} A \emph{mixed
graph} is a graph containing three types of edges denoted by
arrows, arcs (bi-directed edges), and lines (full lines). Notice that
we allow
multiple edges of the same type.
A \emph{loopless mixed graph} (LMG) is a mixed graph that does not
contain any loops (a loop may be line, arrow, or arc).
For an arrow $j\fra i$, we say that the arrow is \emph{from} $j$ \emph
{to} $i$. We also call $j$ a \emph{parent} of $i$, $i$~a \emph{child}
of $j$ and we use the notation $\pa(i)$ for the set of all parents of
$i$ in the graph.
In the cases of $i\fra j$ or $i\arc j$, we say that there is \emph{an
arrowhead at} $j$ or \emph{pointing to} $j$.

A path $\langle i=i_0,i_1,\dots,i_n=j\rangle$ is \emph
{direction-preserving} from $i$ to $j$ if all $i_ki_{k+1}$ edges are
arrows pointing from $i_k$ to $i_{k+1}$. If there is a
direction-preserving path from $j$ to $i$ then $j$ is an \emph
{ancestor} of $i$ and $i$ is a \emph{descendant} of $j$. We denote the
set of ancestors of $i$ by $\an(i)$. Notice that we do not include $i$
in its set of anteriors or descendants.

A \emph{tripath} is a path with three nodes. Note that \cite{sad12}
used the term V-configuration for such a path. However, here we follow
\cite{kii84} and most texts by letting a V-configuration be a tripath
with non-adjacent endpoints.

In a mixed graph the inner node of three tripaths $i\fra
t\fla j$, $i\arc t\fla j$, and \mbox{$i\arc t\arc j$} is
a \emph{collider} (or a collider node) and the inner node of any other tripath
is a \emph{non-collider} (or a non-collider node) on the tripath or
more generally on any path of which the tripath is a subpath. We shall
also say that the tripath itself with inner collider or non-collider
node is a \emph{collider} or \emph{non-collider}. We may speak of a
collider or non-collider without
mentioning the relevant tripath or path when this is apparent from the
context. Notice that a node may be a collider on one tripath and a
non-collider on another.\looseness=-1\vadjust{\goodbreak}

Two paths $\pi_1$ and $\pi_2$ (including tripaths or edges) between
$i$ and $j$ are called \emph{endpoint-identical} if there is an
arrowhead pointing to $i$ in $\pi_1$ if and only if there is an
arrowhead pointing to $i$ in $\pi_2$ and similarly for $j$. For
example, the paths $i\fra j$, $i\ful k\arc j$, and $i\fra k\fla l \arc
j$ are all endpoint-identical as they have an arrowhead pointing to $j$
but no arrowhead pointing to $i$ on the paths.\vspace*{-2pt}

%
%f1 #&#
\begin{figure}[b]\vspace*{-2pt}

\includegraphics{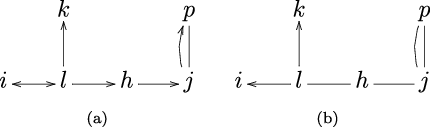}

\caption{(a) A mixed graph $G$. (b) The anterior graph $G^*$ of $G$.}
\label{fig2ex100n}
\end{figure}

%s3.2 #&#
\subsection{Anterior graphs and sets}\vspace*{-2pt}
The \emph{anterior graph} of a
loopless mixed graph $G$, denoted by $G^*$, is the graph obtained from
$G$ by recursively removing arrowheads pointing to nodes that are the
endpoints of a line, that is, by obtaining $\ful\circ\ful$ and $\fla
\circ\ful$ from $\fra\circ\ful$ and $\arc\circ\ful$
respectively. Hence, it holds that $G=G^*$ if and only if there are no
arrowheads pointing to lines in $G$. Notice also that since removing an
arrowhead pointing to a line does not affect other arrowheads pointing
to lines, it does not matter which arrowhead is removed first;
therefore, the order of removing arrowheads pointing to lines does not
affect the final graph obtained.

A path $\langle i=i_0,i_1,\dots,i_n=j\rangle$ from $i$ to $j$ ($i\neq
j$) in $G^*$ is an \emph{anterior path} if it has the form $i\ful
i_1\ful\cdots\ful i_m\fra i_{m+1}\fra\cdots\fra j$. Notice
that this path may only contain lines or arrows. We shall say that $i$
is \emph{anterior} of $j$ in $G$ if there is an anterior path from $i$
to $j$ in $G^*$. Notice that although the anterior path is defined in
$G^*$ we may from time to time refer to an anterior path in $G$ as the
path corresponding to the anterior path in $G^*$.

We use the notation $\ant(i)$ for the set of all anteriors of $i$.
Notice that, since ancestral graphs have no arrowheads pointing to
lines, we have $G=G^*$ for an ancestral graph. Thus, our definition of
anterior extends the notion of anterior used in \cite{ric02} for
ancestral graphs with the minor difference that we do not include a
node in its anterior set. However, it is different from and
inconsistent with the definition of anteriors in \cite{fry90}
and~\cite{and97}.\looseness=-1

For example, in the graph $G$ in Figure~\ref{fig2ex100n}(a), $\ant
(i)=\{l,h,j,p\}$ and $\ant(p)=\{l,h,j\}$. This can be seen by looking
at the anterior paths $\langle p,j,h,l,i\rangle$ from $p$ to $i$ and
$\langle l,h,j,p\rangle$ from $l$ to $p$ (as well as from $p$ to $l$)
in Figure~\ref{fig2ex100n}(b).

We first show that transitivity holds for anteriors.
%
%le1 #&#
\begin{lemma}\label{propve}
For any loopless mixed graph it holds that
if $i\in\ant(j)$ and $j\in\ant(k)$ then $i\in\ant(k)$.\vadjust{\goodbreak}
\end{lemma}
\begin{pf}
If $i\in\ant(j)$ and $j\in\ant(k)$, $G^*$ has anterior paths $\pi_1$ from $i$ to $j$ and $\pi_2$ from $j$ to $k$. As no arrowhead meets
a line in $G^*$ their combination $\pi_1\circ\pi_2$ is an anterior
path from $i$ to $j$ in $G^*$.
\end{pf}

Here we also introduce a lemma that is used in several proofs
of this paper.
%
%le2 #&#
\begin{lemma}\label{lemvvn} Let $G$ be a loopless mixed graph.
If $i\in\ant(j)\setminus\an(j)$, then either $i$ or a descendant of
$i$ is the endpoint of a line in $G$.
\end{lemma}
\begin{pf}
The proof uses induction on the number of arrowheads removed from $G$
to obtain $G^*$.
For the base, if $G=G^*$ it follows immediately from the definition of
an anterior path that $i$ must be the endpoint of a line or we would
have $i\in\an(j)$.

Next, suppose that $G^*$ is obtained from $G$ by removing $n+1$
arrowheads and let $\tilde G$ be obtained from $G$ by removing a single
arrowhead pointing to a line from $G$. Then $G^*$ is also the anterior
graph of $\tilde G$, but with only $n$ arrowheads needing removal.
Thus, if $i\in\ant(j)$ in $G$, it is also anterior to $j$ in $\tilde G$.
Consider now two cases:

\emph{Case I.} Assume $i$ is an ancestor of $j$ in $\tilde G$. Since
$i$ is not an ancestor of $j$ in $G$, $\tilde G$ must have been
obtained by turning an arc into an arrow. Say this arrowhead points to
$h$. Then $h$ is an endpoint of a line and it is a descendant of $i$ in $G$.

\emph{Case II.} If $i$ is not an ancestor of $j$ in $\tilde G$, the
inductive hypothesis yields that $i$ is either adjacent to a line $ih$
in $\tilde G$ or has a descendant $h$ in $\tilde G$ which is the
endpoint of a line in $\tilde G$. Let $h$ be the node adjacent to a
line in $\tilde G$. If the arrowhead removed is not on the
direction-preserving path $\pi$ from $i$ to $h$ the conclusion
obviously follows. Else, there must be node $k$ on $\pi$ which is
adjacent to a line in $G$ and can be used instead of $h$.
\end{pf}

%s3.3 #&#
\subsection{The $m$-separation criterion}
Here we define a separation
criterion for LMGs. We use this criterion to induce independencies on
LMGs and its subclasses defined in Section~\ref{seclmg}.

We first define an $m$-connecting path:
Let $C$ be a subset of the node set of an LMG. A path is \emph
{$m$-connecting} given $C$ if all its
collider nodes are in $C\cup\an(C)$ and all its non-collider nodes
are outside $C$. For two disjoint subsets of the node set $A$ and $B$,
we say that $C$ $m$-separates $A$ and $B$ if there is no $m$-connecting
path between $A$ and $B$ given $C$. In this case, we use the notation
$A\perp_mB\cd C$. Notice that the $m$-separation criterion induces an
independence model $\mathcal{J}_m(G)$ on $G$ by $A\perp_m B\cd C \iff
\langle A,B\cd C\rangle\in\mathcal{J}_m(G)$.

We note that $m$-separation is unaffected if we replace multiple edges
of the same type with a single edge of that type. The $m$-separation
criterion for LMGs is the same as the separation criterion defined in
\cite{ric02}. It is an extension of the $d$-separation criterion
introduced in \cite{pea88}. Clearly, $m$-separation is also an
extension of simple separation in an undirected graph, as then all
edges are lines.

\begin{figure}

\includegraphics{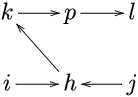}

\caption{A loopless mixed graph $G$ for which $\langle i,j\cd
l\rangle\notin\mathcal{J}_m(G)$.}
\label{figlmg-illustration}
\end{figure}

For example, in graph $G$ in Figure~\ref{figlmg-illustration} it holds
that $h\in\an(l)$ and, thus, $\langle i,h,j\rangle$ is an
$m$-connecting path given $l$. Therefore, $\langle i,j\cd l\rangle\notin\mathcal{J}_m(G)$.
%
%f2 #&#
We now have the following theorem. A~similar result for the induced
independence model for MC graphs was given in Proposition~2.10 of
\cite{kos02}.
%
%th1 #&#
\begin{theorem}\label{thm110}
For any loopless mixed graph $G$, the independence model $\mathcal
{J}_m(G)$ is a compositional graphoid.
\end{theorem}
\begin{pf}
For $G=(N,F)$ and disjoint subsets $A$, $B$, $C$, and $D$ of $N$, we
prove that $\perp_m$ satisfies the six compositional graphoid axioms:
\begin{longlist}[(6)]
\item[(1)] \emph{Symmetry}: If $A\perp_mB\cd C$, then $B\perp_mA\cd C$: If
there is no $m$-connecting path between $A$ and $B$ given $C$, then
there is no $m$-connecting path between $B$ and $A$ given~$C$.

\item[(2)] \emph{Decomposition}: If $A\perp_m(B\cup D)\cd C$, then $A\perp_mD\cd C$: If there is no $m$-connecting path between $A$ and $B\cup D$
given $C$, then there is no $m$-connecting path between $A$ and
$D\subseteq(B\cup D)$ given $C$.

\item[(3)] \emph{Weak union}: If $A\perp_m(B\cup D)\cd C$ then $A\perp_mB\cd
(C\cup D)$: From (2) we know that $A\perp_mD\cd C$ and $A\perp_mB\cd
C$. Suppose, for contradiction, that there exist $m$-connecting paths
between $A$ and $B$ given $C\cup D$. Consider a shortest path of this
type and call it $\pi$. If there is no inner collider node on $\pi$,
then there is an $m$-connecting path between $A$ and $B$ given $C$, a
contradiction. On $\pi$ all collider nodes are in $(C\cup D)\cup\an
(C\cup D)$. If all collider nodes are in $C\cup\an(C)$, then there is
an $m$-connecting path between $A$ and $B$ given $C$, again a
contradiction. Hence, consider the closest collider node $i\in(D\cup
\an(D))\setminus(C\cup\an(C))$ to $A$ on $\pi$. Now since the
nodes between $A$ and $i$ are not in $B\cup D$, there is an
$m$-connecting path between $A$ and $i$ given $C$. If $i\in D$, then
this is obviously a contradiction. Otherwise there is a node $k\in D$,
for which $i\in\an(k)$ and thus an $m$-connecting path between $A$
and $k$ given $C$, a contradiction again. Therefore, there is no
$m$-connecting path between $A$ and $B$ given $C\cup D$.

\item[(4)] \emph{Contraction}: If $A\perp_mB\cd C$ and $A\perp_mD\cd (B\cup
C)$, then $A\perp_m(B\cup D)\cd C$: Suppose, for contradiction, that
there exists an $m$-connecting path between $A$ and $B\cup D$ given
$C$. Consider a shortest path of this type and call it $\pi$. The path
$\pi$ is either between $A$ and $B$ or between $A$ and $D$. The path
$\pi$ being between $A$ and $B$ contradicts $A\perp_mB\cd C$.
Therefore, $\pi$ is between $A$ and $D$. In addition, since all inner
collider nodes on $\pi$ are in $C\cup\an(C)$ and because $A\perp_mD\cd (B\cup C)$, an inner non-collider node should be in $B$. This
contradicts the fact that $\pi$ is a shortest $m$-connecting path
between $A$ and $B\cup D$ given $C$.

\item[(5)] \emph{Intersection}: If $A\perp_mB\cd (C\cup D)$ and $A\perp_mD\cd
(C\cup B)$, then $A\perp_m(B\cup D)\cd C$: Suppose, for contradiction,
that there exists an $m$-connecting path between $A$ and $B\cup D$
given $C$. Consider a shortest path of this type and call it $\pi$.
The path $\pi$ is either between $A$ and $B$ or between $A$ and $D$.
Because of symmetry between $B$ and $D$ in the formulation it is enough
to suppose that $\pi$ is between $A$ and $B$. Since all inner collider
nodes on $\pi$ are in $C\cup\an(C)$ and because $A\perp_mB\cd (C\cup
D)$, an inner non-collider node should be in $D$. This contradicts the
fact that $\pi$ is a shortest $m$-connecting path between $A$ and
$B\cup D$ given $C$.

\item[(6)] \emph{Composition}: If $A\perp_mB\cd C$ and $A\perp_mD\cd C$, then
$A\perp_m(B\cup D)\cd C$: Suppose, for contradiction, that there exist
$m$-connecting paths between $A$ and $B\cup D$ given $C$. Consider a
path of this type and call it $\pi$. Path $\pi$ is either between $A$
and $B$ or between $A$ and $D$. Because of symmetry between $B$ and $D$
in the formula it is enough to suppose that $\pi$ is between $A$ and
$B$. But this contradicts $A\perp_mB\cd C$.\qed
\end{longlist}\noqed
\end{pf}

Theorem~\ref{thm110} implies that we can focus on establishing
conditional independence for pairs of nodes, formulated in the
corollary below.
%
%co1 #&#
\begin{coro}
For a loopless mixed graph $G$ and disjoint subsets of its node set
$A$, $B$, and $C$, it holds that $A\perp_mB\cd C$ if and only if $i\perp_m j\cd C$ for every nodes $i\in A$ and $j\in B$.
\end{coro}
\begin{pf}
The result follows from the fact that $\perp_m$ satisfies the
decomposition and the composition properties.
\end{pf}

%s4 #&#
\section{Subclasses of loopless mixed graphs}
\label{secspecial}

LMGs and their associated independence models induced by $m$-separation
unify a variety of previously discussed graphical independence models.
%s4.1 #&#
\subsection{Chain graphs}
Important exceptions include certain independence models for chain
graphs. Chain graphs themselves are LMGs, but at least four different
Markov properties for chain graphs have been discussed in the
literature. Drton \cite{drt09} has classified them into (i) the \emph
{LWF} or \emph{block concentration} Markov property, (ii) the \emph{AMP}
or \emph{concentration regression} Markov property, (iii) a Markov
property that is dual to the AMP Markov property, and (iv) and the
\emph{multivariate regression}
Markov property. When the chain components consist entirely of arcs,
the multivariate regression property is identical to the one induced by
$m$-separation. However, the independence model induced by
$m$-separation in a chain graph is typically different from any of the
other chain graph interpretations; see also \cite
{richardson98,ric01} and \cite{lauritzenrichardson02}.
%s4.2 #&#
\subsection{Ribbonless graphs}
The class of MC graphs, defined in \cite{kos02}, contains line loops
and uses a different separation criterion for inducing an independence
model. However, a small modification of any MC graph that is derived
from a DAG after marginalisation and conditioning yields a so-called
ribbonless graph, which is loopless and induces the same independence
model as the MC graph, but by $m$-separation \cite{sadthesis}. Any
ribbonless graph can be generated from a DAG by marginalisation and
conditioning and ribbonless graphs are stable under these operations
\cite{sad12}.
The remaining part of this paper deals with such graphs. We first give
a formal definition of a ribbon.

A \emph{ribbon} is a collider tripath $\langle h,i,j\rangle$ such
that both of the following two conditions hold:
\begin{enumerate}[(2)]
\item there is no endpoint-identical edge between $h$ and $j$, that is,
there is no $hj$-arc in the case of $h\arc i\arc j$; there is no
$hj$-line in the case of $h\fra i\fla j$; and there is no arrow from
$h$ to $j$ in the case of $h\fra i\arc j$;
\item$i$ or a descendant of $i$ is the endpoint of a line or is on a
direction-preserving cycle.
\end{enumerate}

\begin{figure}

\includegraphics{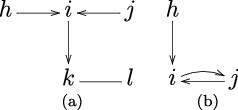}

\caption[]{(a) A straight ribbon $\langle h,i,j\rangle$ with $\nei
(i)=\varnothing$. (b) The simplest cyclic ribbon $\langle h,i,j\rangle$.}
\label{fig3ex2}
\end{figure}

If $i$ or a descendant of $i$ is the endpoint of a line, then we say
the ribbon is \emph{straight} and if they are on a
direction-preserving cycle we say the ribbon is \emph{cyclic}.
A \emph{ribbonless graph} (RG) is an LMG that has no ribbons as
induced subgraphs.
Figure~\ref{fig3ex2} illustrates a straight ribbon $\langle
h,i,j\rangle$ and the simplest cyclic ribbon.
%
%f3 #&#

\begin{figure}[b]

\includegraphics{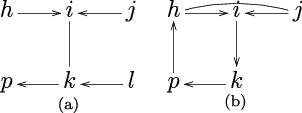}

\caption{(a) A graph that is not ribbonless. (b) A ribbonless graph.}
\label{fig3ex2nn}
\end{figure}

Figure~\ref{fig3ex2nn}(a) illustrates a graph containing a straight
ribbon $\langle h,i,j\rangle$ and Figure~\ref{fig3ex2nn}(b)
illustrates a ribbonless graph. Notice that $\langle h,i,j\rangle$ is
not a ribbon here since there is a line between $h$ and $j$ and this is
an endpoint-identical edge.
%
%f4 #&#
We proceed to establish that ribbonless graphs yield identical
independence models to their anterior graphs and need the following lemma.
%
%le3 #&#
\begin{lemma}\label{lemvvnn}
Let $G$ be a ribbonless graph.
If there is a collider tripath $\langle i,j,k\rangle$ in $G$ that is
non-collider in $G^*$, then $G$ has an $ik$-edge that is
endpoint-identical to $\langle i,j,k\rangle$.
\end{lemma}
\begin{pf}
Suppose that $\langle G=G_0,G_1,\dots,G_n=G^*\rangle$ is a sequence
of graphs, where each graph has been generated by removing one
arrowhead pointing to a full line from the previous graph starting from $G$.

Consider the first intermediate graph $G_{p+1}$ where $\langle
i,j,k\rangle$ turns into a non-collider tripath. We prove by reverse
induction that, for each $0\leq q \leq p$, $\langle i,j,k\rangle$ is a
straight ribbon unless there is an endpoint-identical $ik$-edge to
$\langle i,j,k\rangle$.

In $G_p$, the node $j$ is obviously the endpoint of a line and the
result holds. Thus, we assume that the result holds for $G_{q}$. In
$G_{q-1}$, it is easy to observe that if the line that makes the ribbon
is an arrow pointing to another line or if an arrow on the
direction-preserving cycle pointing to a line is an arc then $j$ or a
descendant of $j$ is still the endpoint of a line. Therefore, the
result holds in $G_{q-1}$. Therefore, by reverse induction, this result
holds in $G$, and since $G$ is ribbonless, in $G$ there is an
endpoint-identical $ik$-edge to $\langle i,j,k\rangle$.
\end{pf}
For the graph $G$ in Figure~\ref{fig3ex2}(a), the anterior graph
$G^*$ is the graph where all edges become undirected. Clearly there is
no endpoint-identical edge $hj$ and the conclusion of Lemma~\ref
{lemvvnn} does not hold. This illustrates the role of a graph being ribbonless.
%
%pr1 #&#
\begin{prop}\label{propvvn}
For a ribbonless graph $G$, it holds that $\mathcal{J}_m(G)=\mathcal
{J}_m(G^*)$, that is, $G$ and $G^*$ are Markov equivalent.
\end{prop}
\begin{pf}
It is enough to prove that there is an $m$-connecting path between $i$
and $j$ given $C$ in $G$ if and only if there is an $m$-connecting path
between $i$ and $j$ given $C$ in $G^*$.

Suppose that there is an $m$-connecting path between $i$ and $j$ given
$C$ in $G$. All non-colliders on the path in $G$ are preserved in
$G^*$. In addition, by Lemma~\ref{lemvvnn}, a collider tripath
$\langle i,j,k\rangle$ becomes non-collider if there is an
endpoint-identical $ik$-edge to $\langle i,j,k\rangle$. In this case,
the $ik$-edge can be used instead of $\langle i,j,k\rangle$ to
establish an $m$-connecting path in~$G^*$.\looseness=1

Conversely, suppose that there is an $m$-connecting path between $i$
and $j$ given $C$ in $G^*$. Collider tripaths are collider tripaths in
$G$, and if a non-collider tripath $\langle i,j,k\rangle$ has been
collider in $G$ then, by Lemma~\ref{lemvvnn}, one can again use the
$ik$-edge instead of $\langle i,j,k\rangle$. Thus the only thing that
remains to be proven is that a direction-preserving path pointing to a
member of $C$ in $G$ remains direction-preserving in $G^*$.

In this case, by the same argument as in Lemma~\ref{lemvvnn}, if for
the collider tripath $\langle i,j,k\rangle$, where $j\in\an(C)$, the
arrowhead of an arrow on the direction-preserving path in $G$ is taken
away then $\langle i,j,k\rangle$ is a ribbon unless there is an
endpoint-identical $ik$-edge to $\langle i,j,k\rangle$. Hence, we can
use the $ik$-edge instead of $\langle i,j,k\rangle$ to establish an
$m$-connecting path.
\end{pf}

Thus, the absence of ribbons ensures that the Markov property is
unchanged by forming the anterior graph $G^*$. Again, as the anterior
graph $G^*$ of the graph $G$ in Figure~\ref{fig3ex2}(a) is the graph
with all edges becoming undirected, we have $h\perp_m j$ in $G$ but not
$h\perp_m j$ in $G^*$, illustrating that absence of ribbons is
essential for the Markov equivalence of $G$ and~$G^*$.

Independence models induced by $m$-separation in a ribbonless graph can
be induced by marginalisation over and conditioning on a
DAG-independence model \cite{sad12}. This implies that independence
models corresponding to RGs are probabilistic, that is, any RG has a
faithful probability distribution.
%s4.3 #&#
\subsection{Other subclasses of loopless mixed graphs}
Other subclasses of LMGs that use $m$-separation and have been
discussed in the literature are \emph{summary graphs} \cite{wer11},
\emph{ancestral graphs} \cite{ric02}, \emph{acyclic directed mixed
graphs} \cite{spi97,ric03},
\emph{undirected} or \emph{concentration graphs} \cite
{dar80,lau96}, \emph{bidirected} or \emph{covariance graphs} \cite
{cox93,kau96,wer98,drtonrichardson08}, and the class of directed
acyclic graphs \cite{kii84,pea88,gei90}. In papers on summary graphs
and regression chain graphs, dashed undirected
edges (without arrowheads) have often been used in place of bi-directed
edges. Using the latter as we have done here makes the idea of a
collider more immediate so $m$-separation can be used directly and the
relation between the various types of graphs becomes transparent.

The use of some of the above graphs are motivated by representing
independence models obtained by marginalisation over and conditioning
on subsets of the node set of a DAG. For those graphs, arcs indicate
marginalisation and lines indicate conditioning.

\begin{figure}

\includegraphics{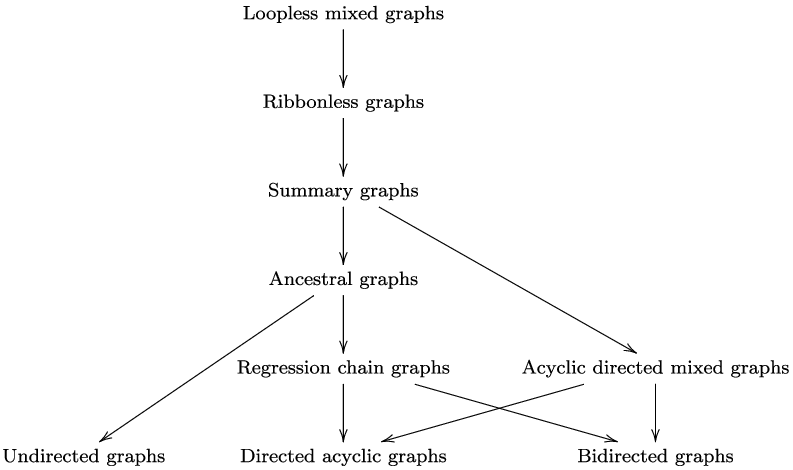}

\caption{The hierarchy of subclasses of LMGs.}
\label{fig1110}
\end{figure}

The diagram in Figure~\ref{fig1110} illustrates the hierarchy of
subclasses of LMGs and their associated independence models generated
by $m$-separation. For example, it can be seen from the diagram that
bidirected graphs are also ancestral graphs, since they form a subclass
of multivariate regression chain graphs, which again form a subclass of
ancestral graphs. Notice that the associated classes of independence
models are all distinct except for ancestral, summary, and ribbonless
graphs, which are alternative representations of the same class of
independence models.
%
%f5 #&#

%s5 #&#
\section{Maximal ribbonless graphs}
\label{secmaximal}

Among the independence models over the node set $V$ of a graph $G$,
those that are of interest to us \emph{conform} with $G$, meaning that
$i\sim j$ in $G$ implies $\langle i,j\cd C\rangle\notin\mathcal{J}$
for any $C\subseteq V\setminus\{i,j\}$. Henceforth, we assume that
independence models $\mathcal{J}$ conform with $G$, unless otherwise stated.

\begin{figure}[b]

\includegraphics{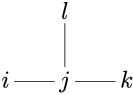}

\caption{The independence model $\mathcal{J}=\{\langle i,l\cd
j\rangle,\langle i,k\cd \varnothing\rangle\}$ conforms with $G$
whereas $\mathcal{J}=\{\langle i,l\cd j\rangle,\langle i,j\cd
\varnothing\rangle\}$ does not.}
\label{fig1nn}
\end{figure}

For example, the independence model $\mathcal{J}=\{\langle i,l\cd
j\rangle,\langle i,k\cd \varnothing\rangle\}$ conforms with the
graph $G$ in Figure~\ref{fig1nn}, whereas $\mathcal{J}=\{\langle
i,l\cd j\rangle,\langle i,j\cd \varnothing\rangle\}$ does not
conform with $G$ because of the independence statement $\langle i,j\cd
\varnothing\rangle$.
%
%f6 #&#

A ribbonless graph $G$ is called \emph{maximal} if by adding any edge
to $G$, the independence model induced by $m$-separation changes.
Note that in \cite{wer11} a graph that is maximal is called an \emph
{independence graph}.

The independence models on RGs induced by $m$-separation conform with
the graphs; hence for maximal graphs, adding an edge to the graph makes
the independence model smaller. Therefore, we have the lemma below.
%
%le4 #&#
\begin{lemma}\label{lem111}
A graph $G=(V,E)$ is maximal if and only if for every pair of
non-adjacent nodes $i$ and $j$ of $V$, there exists a subset $C$ of
$V\setminus\{i,j\}$ such that $i\perp_m j\cd C$.
\end{lemma}
\begin{pf}
The result follows directly from the definition of maximality.
\end{pf}

\begin{figure}

\includegraphics{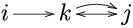}

\caption{A non-maximal RG.}
\label{fig11ex6}
\end{figure}

RGs are not maximal in general. To see this consider the RG in Figure~\ref{fig11ex6}. There is no $C$ such that $i\perp_m j\cd C$. This is
because if $k\in C$, the path $i\fra k \arc j$ is $m$-connecting given
$C$, and if $k\notin C$, $i\fra k \fra j$ is $m$-connecting given $C$.
%
%f7 #&#

To characterise maximal RGs, we need the following notion:
A path $\langle j,q_1,q_2,\dots,q_p,i\rangle$ is a \emph{primitive
inducing} path between $i$ and $j$ if and only if for every $n$,
$1\leq n \leq p$,
\begin{enumerate}[(ii)]
\item[(i)] $q_n$ is a collider on the path; and
\item[(ii)] $q_n\in\an(\{i\}\cup\{j\})$.
\end{enumerate}
This definition is a trivial extension of a primitive inducing path as
defined for ancestral graphs in \cite{ric02}.
Note in particular that we consider any edge between $i$ and $j$ to be
a primitive inducing path. In Figure~\ref{fig11ex6}, $\langle i, k,
j\rangle$ is a primitive inducing path.

Next, we need the following lemmas. These also establish a pairwise
Markov property for maximal RGs.
%
%le5 #&#
\begin{lemma}\label{lem11300}
A non-collider node $k$ on a path $\pi$ between $i$ and $j$ in a
ribbonless graph $G$ is either in $\ant(i)\cup\ant(j)$ or an
anterior of a collider node $h$ on $\pi$. Moreover, the relevant
subpath of $\pi$ between $k$ and $i$, $j$ or $h$ is an anterior path
in $G^*$.
\end{lemma}
\begin{pf} Let $k=i_m$ be a non-collider node on
a path $\pi=\langle i=i_0,i_1,\dots,i_n=j\rangle$ in $G$. Then from
at least one side
(say from $i_{m-1}$) there is no arrowhead on $\pi$ pointing to $k$.
By moving towards $i$ on the path as long as $i_p$, $1\leq p\leq m-1$,
is non-collider on the path, we obtain that $k\in\ant(i_{p-1})$.
This implies that if no $i_p$ is a collider then $k\in\ant(i)$ and
hence the lemma follows.
\end{pf}
%
%le6 #&#
\begin{lemma}\label{lem1130}
For nodes $i$ and $j$ in an RG that are not connected by any primitive
inducing paths (and hence $i\not\sim j$), it holds that $i\perp_m j\cd
(\ant(i)\cup\ant(j))\setminus\{i,j\}$.
\end{lemma}
\begin{pf}
Suppose, for contradiction, there is an $m$-connecting path between $i$
and $j$ given $(\ant(i)\cup\ant(j))\setminus\{i,j\}$ and denote a
shortest such path by $\pi$.
If there is a non-collider node $k$ on $\pi$ then, by Lemma~\ref
{lem11300}, $k$ is either in $\ant(i)\cup\ant(j)$ or it is an
anterior of a
collider node on $\pi$. But since $\pi$ is $m$-connecting given
$(\ant(i)\cup\ant(j))\setminus\{i,j\}$, collider nodes are in $\ant
(i)\cup\ant(j)$ themselves. Hence, $k\in\ant(i)\cup\ant(j)$,
which contradicts the fact that $\pi$ is $m$-connecting. Therefore,
all inner nodes of $\pi$ must be colliders.

Now we know that all inner nodes of $\pi$ are in $\ant(i)\cup\ant
(j)$ and $i\not\sim j$. If, for a collider tripath $\langle
r,l,s\rangle$ on $\pi$, $l\in(\ant(i)\cup\ant(j))\setminus(\an
(i)\cup\an(j))$ then, by Lemma~\ref{lemvvn} and since the graph is
ribbonless, there is an endpoint-identical $rs$-edge to the tripath,
which contradicts $\pi$ being shortest. Therefore, $l\in\an(i)\cup
\an(j)$, which implies that $\pi$ is primitive inducing, again a
contradiction. Therefore, there is no $m$-connecting path between $i$
and $j$ given $(\ant(i)\cup\ant(j))\setminus\{i,j\}$, and hence
$i\perp_m j\cd (\ant(i)\cup\ant(j))\setminus\{i,j\}$.
\end{pf}
Next, in Theorem~\ref{thm112} we give a necessary and sufficient
condition for an RG to be maximal. The analogous result for ancestral
graphs was proved in Theorem~4.2 of \cite{ric02}.
%
%th2 #&#
\begin{theorem}\label{thm112}
A ribbonless graph $G$ is maximal if and only if $G$ does not contain
any primitive inducing paths between non-adjacent nodes.
\end{theorem}
\begin{pf}
Let $\pi=\langle i=i_0,i_1,\dots,i_n=j\rangle$ be a primitive
inducing path between $i$ and $j$ in
$H$, and let $C$ be a subset $V\setminus\{i,j\}$, where $V$ is the
node set of $H$. We need to show that there is an $m$-connecting path
between $i$ and $j$ given $C$.

This is immediate if each internal node, that is, each of $i_1,\dots
,i_{n-1}$, is in $C\cup\an(C)$ by
just using $\pi$, so assume that this is not the case. Thus there is
an internal node
of $\pi$ not in $C\cup\an(C)$, and we may assume that there is one
in $\an(i)$. Pick
such a node $i_q$, $1\leq q< n$, as far along the path to $j$ as
possible. Consider a
direction-preserving path from $i_q$ to $i$, and let $P_1$ denote the
reverse of this path.
Note that no internal node in $P_1$ is in $C\cup\an(C)$. Let $\pi_1$
be the part of $\pi$ from $i_q$ to
$j$. If each internal node in this path is in $C\cup\an(C)$ then we
are done by taking the
path $P_1$ followed by $\pi_1$ (note that no node can be repeated
since each internal node
in $\pi_1$ is in $C\cup\an(C)$ and each internal node in $P_1$ is
outside $C\cup\an(C)$). So suppose
not. Let $i_p$ be the first node in $\pi_1$ that is not in $C\cup\an
(C)$. Then $i_p\notin\an(i)$ (by the way $i_q$ was chosen), so
$i_p\in\an(j)$. Let $\pi_2$ be the part of $\pi$ from $i_q$ to
$i_p$, and let
$P_2$ be a direction-preserving path from $i_p$ to $j$. Note that no
internal node in $P_2$
is in $C\cup\an(C)$. If $P_1$ and $P_2$ have no intersection, then
much as above we obtain
an $m$-connecting path given $C$ by taking $P_1$ followed by $\pi_2$,
followed by $P_2$. If $P_1$
and $P_2$ do intersect, then we obtain an $m$-connecting path as
required by
following $P_1$ up to the first node on $P_2$ and then following $P_2$.
%If there exists a pair of non-adjacent nodes between $i$ and $j$
%connected by a primitive inducing path $\pi$ then regardless of
%whether the inner nodes are
%in $C\cup\an(C)$ or not, an $m$-connecting path between $i$ and $j$
%can be found. Therefore, $G$ is non-maximal. More precisely, suppose
%that
%$\pi=\langle i=i_0,i_1,\dots,i_n=j\rangle$. We trace an $m$-connecting
%path between $i$ and $j$ by the following method. We start from $i$.
%We consider the closest node $k=i_q$ to $j$ on $\pi$ such that it is
%in $\an(i)$ and the nodes on the direction-preserving path are not in
%$C$. We proceed to $k$ through the direction-preserving path. Hence
%all $i_p$, $q\leq p\leq n-1$ are either in $\an(C)$ or in $\an(j)
%in $\an(j)\setminus\an(C)$ then we proceed to $j$ through the nodes of
%the direction-preserving path.

By letting $C=(\ant(i)\cup\ant(j))\setminus\{i,j\}$ for every
non-adjacent nodes $i$ and $j$, the other direction follows from Lemmas
\ref{lem111} and \ref{lem1130}.
\end{pf}
For other special types of graphs that are subclasses of RGs, the
condition for maximality of RGs may get further simplified. Among the
subclasses of RGs that have been mentioned in this paper, summary
graphs, ancestral graphs, and acyclic directed mixed graphs are not
necessarily maximal, while all others are maximal. This can be seen by
checking whether primitive inducing paths are permissible in each subclass.

A Markov equivalent maximal graph can be generated from a non-maximal
graph by adding endpoint-identical edges to a primitive inducing path
between a pair of non-adjacent nodes. We refer the reader to \cite
{sadthesis} for details. The following lemma establishes that anterior
graphs of maximal graphs are themselves maximal.
%
%le7 #&#
\begin{lemma}\label{lemantmax} Let $G$ be a ribbonless graph and
$G^*$ its anterior graph. Then if $G$ is maximal, so is $G^*$.
\end{lemma}
\begin{pf}
If, for contradiction, $G^*$ is not maximal, then Theorem~\ref
{thm112} implies that there is a primitive inducing path in $G^*$
between non-adjacent nodes $i$ and $j$. Consider a shortest primitive
inducing path between $i$ and $j$ and denote it by $\pi$. We know that
all inner nodes of $\pi$ are colliders in $G^*$. This trivially
implies that all inner nodes of $\pi$ are colliders in $G$ too. In
addition, each inner node $k$ on $\pi$ is in $\an(\{i,j\})$ in $G^*$.
In $G$, $k\in\an(\{i,j\})$ unless an arrow on the
direction-preserving path from $k$ to $i$ or $j$ is an arc turning into
an arrow in $G^*$. In this case, $k$ is an ancestor of a node that is
the endpoint of a line. Hence the tripath $\langle h,k,l\rangle$ on
$\pi$ is a ribbon unless there is an endpoint-identical $hl$-edge to
the tripath, which contradicts the fact that $\pi$ is shortest.
Therefore, $\pi$ is a primitive inducing path in $G$, a contradiction.
Hence, $G^*$ is maximal.
\end{pf}

%s6 #&#
\section{Markov properties for ribbonless graphs}
\label{secmarkov}
In this section, we give a precise definition of the global and
pairwise Markov properties for an independence model $\mathcal{J}$
defined over the node set of a ribbonless graph. Further we show that
these two Markov properties are equivalent for a maximal ribbonless
graph if $\mathcal{J}$ is also a compositional graphoid. This result
is a direct generalisation of the similar result of \cite{pea88} for
undirected graphs and graphoids.

%s6.1 #&#
\subsection{Global and pairwise Markov properties}

For a ribbonless graph $G=(V,E)$, an independence model $\mathcal{J}$
defined over $V$ satisfies the \emph{global Markov property} w.r.t.
$G$ if it holds for $A$, $B$, and $C$ disjoint subsets of $V$ that
\[
A\perp_{m} B\cd C \quad \Longrightarrow\quad \langle A,B\cd C\rangle
\in\mathcal{J}.
\]

Similarly, an independence model $\mathcal{J}$ defined over $V$
satisfies the \emph{pairwise Markov property} w.r.t. $G$ if it holds
for any nodes $i$ and $j$ that
\[
i\not\sim j\quad \Longrightarrow\quad \bigl\langle i,j\cd \bigl(\ant(i)\cup
\ant(j) \bigr)\setminus\{ i,j\} \bigr\rangle\in\mathcal{J}.
\]

\begin{figure}[b]

\includegraphics{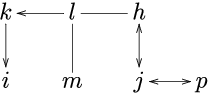}

\caption{The pairwise Markov property for this RG implies, for
example, $\langle i,m\cd \{k,l,h\}\rangle$. The global Markov property
would for example imply $\langle\{i,k\}, j \cd l\rangle$.}
\label{figpairglobal}
\end{figure}

For example, for the graph in Figure~\ref{figpairglobal}, the pairwise
Markov property would imply that $\langle i,m\cd \{k,l,h\}\rangle$ as
$\ant(i) =\{k,l,h,m\}$ and $\ant(m)= \{l,h\}$. It would also imply
that $\langle l,p\cd \{h,m\}\rangle$.
%
%f8 #&#

Clearly, the independence model $\mathcal{J}_m(G)$ induced by
$m$-separation always satisfies the global Markov property w.r.t. $G$.
By Lemma~\ref{lem111}, Lemma~\ref{lem1130}, and Theorem~\ref
{thm112}, \emph{$\mathcal{J}_m(G)$ satisfies the pairwise Markov
property if and only if $G$ is maximal}.

%s6.2 #&#
\subsection{Equivalence of pairwise and global Markov properties}
Before establishing the main result of this section, we need two lemmas.
%
%le8 #&#
\begin{lemma}\label{lemj1}
Let $\mathcal{J}$ be a compositional graphoid over a set $V$ and $M$
and $C$ be disjoint subsets of $V$. It then holds that the marginal
independence model
\[
\alpha(\mathcal{J},M)= \bigl\{\langle A,B\cd C\rangle\dvt \langle A,B\cd C
\rangle \in\mathcal{J}\mbox{ and } (A\cup B\cup C)\cap M=\varnothing \bigr\},
\]
which is defined over $V\setminus M$, is a compositional graphoid.
\end{lemma}
\begin{pf}
All the six compositional graphoid properties for $\alpha(\mathcal
{J},M)$ follow trivially from the facts that for $A$, $B$, and $C$ such that
$(A\cup B\cup C)\cap M=\varnothing$, $\langle A,B\cd C\rangle\in
\alpha(\mathcal{J},M)$ if and only if $\langle A,B\cd C\rangle\in
\mathcal{J}$,
and $\mathcal{J}$ satisfies the six properties.
\end{pf}
Notice that the notion of a marginal independence model $\alpha
(\mathcal{J},M)$ is identical to the notion formally defined in \cite
{ric02} with a different notation; it was also discussed in \cite
{sad12} with the same notation as in this paper.

The following lemma gives sufficient conditions for the combination of
two $m$-connecting paths in anterior graphs to be $m$-connecting.
%
%le9 #&#
\begin{lemma}\label{lemj2} Let $G^*$ be the anterior graph of a
ribbonless graph $G$ and suppose that there are paths $\pi_1=\langle
i=i_0,i_1,\dots,i_n,h\rangle$ between $i$ and $h$ and $\pi_2=\langle
h,j_m,j_{m-1},\dots,j_0=j\rangle$
between $h$ and $j$ which are $m$-connecting given $C$. The combination
$\pi_{12}=\pi_1\circ\pi_2$ is then an $m$-connecting path between
$i$ and $j$ given $C$ in each of the following mutually exclusive situations:
\begin{enumerate}[(b2)]
\item[(a1)] $\langle i_n,h,j_m\rangle$ is a collider and $h\in C\cup
\an(C)$;
\item[(a2)] $i_n=j_m$ with an arrowhead pointing to $h$ on the
$i_nh$-edge and $h\in C\cup\an(C)$;
\item[(b1)] $\langle i_n,h,j_m\rangle$ is a non-collider and $h\notin C$;
\item[(b2)] $i_n=j_m$ with no arrowhead pointing to $h$ on the $i_nh$-edge.
\end{enumerate}
\end{lemma}
\begin{pf}
Let $\pi_{12}=\pi_1\circ\pi_2=\langle i,\ldots i_{p-1},
k,j_{q-1},\dots,j\rangle$ be the combination of $\pi_1$ and $\pi_2$.
If $k=h$ and either (a1) or (b1) holds then the conclusion is obvious.
The cases (a2) or (b2) are only relevant when $k\neq h$.

Next consider the situation where $k\neq h$. Since $\pi_1$ and $\pi_2$ are $m$-connecting, for $\pi_{12}$ to be $m$-connecting we only
need to check the
tripath $\langle i_{p-1},k,j_{q-1}\rangle$. We have to deal with two cases:

\textit{Case 1:} $\langle i_{p-1},k,j_{q-1}\rangle$ \textit{is a non-collider}.

In this case there is no arrowhead pointing to $k$ from at least one of
$i_{p-1}$ or $j_{q-1}$. This means
that $\langle i_{p-1},k,i_{p+1}\rangle$ on $\pi_1$ or $\langle
j_{q-1},k,j_{q+1}\rangle$ on $\pi_2$ is a non-collider, and since
$\pi_1$ and $\pi_2$ were both $m$-connecting we have $k\notin C$.
Hence $\pi_{12}$ is $m$-connecting.

\textit{Case 2:} $\langle i_{p-1},k,j_{q-1}\rangle$ \textit{is a collider}.
We need to consider the following two subcases:

\textit{Case 2.1}. If $\langle i_{p-1},k,j_{q-1}\rangle$ is a
collider and any of $\langle i_{p-1},k,i_{p+1}\rangle$ or $\langle
j_{q-1},k,j_{q+1}\rangle$ is also
a collider then $k\in C\cup\an(C)$ and $\pi_{12}$ is $m$-connecting.

\textit{Case 2.2}. If $\langle i_{p-1},k,j_{q-1}\rangle$ is a
collider but $\langle i_{p-1},k,i_{p+1}\rangle$ and $\langle
j_{q-1},k,j_{q+1}\rangle$
are both non-colliders then by Lemma~\ref{lem11300}, the subpath of
$\pi_1$ from $k$ to a collider node $l_1$ or
to $h$ is an anterior path and similarly for $\pi_2$, $l_2$, and $h$.
However, since $G^*$ is an anterior graph and there are arrowheads
pointing to $k$, these anterior paths must be direction-preserving and
thus $k\in\an(l_1)\cup\an(h)$ and $k\in\an(l_2)\cup\an(h)$. Now
we have the two following further subcases:

\emph{Case 2.2.1: One of the subpaths of $\pi_1,\pi_2$ from $k$ to
$l_1,l_2$ is direction-preserving.} Because $\pi_1$ and $\pi_2$ are
$m$-connecting we must have $l_1$ or $l_2$ in $C\cup\an(C)$. Thus,
$k\in\an(C)$
and $\pi_{12}$ is $m$-connecting.

\emph{Case 2.2.2: Both subpaths of $\pi_1$ and $\pi_2$ from $k$ to
$h$ are direction-preserving.} Then $\langle i_n,h,j_m\rangle$ is
collider or $i_n=j_m$ with an arrowhead pointing to $h$ on the
$i_nh$-edge and (b1) and (b2) are impossible. If (a1) or (a2) holds $\pi_{12}$ is $m$-connecting since then $h\in C\cup\an(C)$.
\end{pf}

We are now ready to establish the main result of this paper.
%
%th3 #&#
\begin{theorem}\label{thm113}
Let $G$ be a maximal ribbonless graph. If an independence model
$\mathcal{J}$ over the node set of $G$ is a compositional graphoid,
then $\mathcal{J}$ satisfies the pairwise Markov property w.r.t. $G$
if and only if it satisfies the global Markov property w.r.t. $G$.
\end{theorem}
\begin{pf}
($\Leftarrow$) If $\mathcal{J}$ is a compositional graphoid and
satisfies the global Markov property it follows from Theorem~\ref
{thm112} and
Lemma~\ref{lem1130} that it satisfies the pairwise Markov property.

($\Rightarrow$) Now suppose that $\mathcal{J}$ satisfies the pairwise
Markov property and compositional graphoid axioms. For subsets $A$, $B$,
and $C$ of the node set of $G$, we should prove that $A\perp_m B\cd C$
implies $\langle A,B\cd C\rangle\in\mathcal{J}$. By composition, it is
sufficient to show this when $A$ and $B$ are singletons, that is, that
$i\perp_m j\cd C$ implies $\langle i,j\cd C\rangle\in\mathcal{J}$.

Further we observe that it is sufficient to establish the result in the
case when $G=G^*$ is itself an anterior graph. Proposition~\ref
{propvvn} gives that $A\perp_m B\cd C$ in $G$, which implies $A\perp_m
B\cd C$ in $G^*$. In addition, by Lemma~\ref{lemantmax}, $G^*$ is a
maximal graph. Moreover, $G$ and $G^*$ have the same anterior sets, and
therefore the same pairwise Markov property. Thus in the following, we
assume that $G=G^*$ is an anterior graph.

We prove
the result in two main parts. In part I, we prove the result for the
case that $C\subseteq\ant(i)\cup\ant(j)$. In part II, we use the
result of part I to establish the general
case.

\textit{Part I}. Suppose that $C\subseteq\ant(i)\cup\ant
(j)$. We use induction on the number of nodes of the graph. The
induction base
for a graph with two nodes is trivial. Thus, suppose that the result
holds for all anterior graphs with fewer than $n$ nodes and assume that $G^*$
has $n$ nodes.

Let $D=\{i\}\cup\{j\}\cup\ant(i)\cup\ant(j)$ and $M=V\setminus D$,
where $V$ is the node set of the graph. First in case I.1 we suppose
that $M\neq\varnothing$, and then in case I.2 we suppose that
$M=\varnothing$.

\emph{Case I.1}. Consider $G^*[D]$ to be the subgraph induced by $D$.
Consider the marginal independence model $\alpha(\mathcal{J},M)=\{
\langle A,B\cd C\rangle\dvt \langle A,B\cd C\rangle\in\mathcal{J}$
and  $(A\cup B\cup C)\cap M=\varnothing\}$
defined over $D$. By Lemma~\ref{lemj1}, $\alpha(\mathcal{J},M)$ is
a compositional graphoid. In addition,
it satisfies the pairwise Markov property: This is because two
non-adjacent nodes $l_1$ and $l_2$ in $G^*[D]$ are non-adjacent in
$G^*$ and by the pairwise Markov property for $\mathcal{J}$,
$\langle l_1,l_2\cd (\ant_{G^*}(l_1)\cup\ant_{G^*}(l_2))\setminus\{
l_1,l_2\}\rangle\in\mathcal{J}$, where $\ant_{G^*}$ is the anterior
set in $G^*$. We know that $\ant_{G^*}(l_1)\cup\ant_{G^*}(l_2)\subseteq D$ and hence $\ant_{G^*}(l_1)\cup\ant_{G^*}(l_2)\cap M=\varnothing$. In addition,
for a node $l$ in $G^*[D]$, $\ant_{G^*}(l)=\ant_{G^*[D]}(l)$.
Therefore, $\langle l_1,l_2\cd (\ant_{G^*[D]}(l_1)\cup\ant_{G^*[D]}(l_2))\setminus\{l_1,l_2\}\rangle\in\alpha(\mathcal{J},M)$.

We also know that $i\perp_m j\cd C$ in $G^*$ implies $i\perp_m j\cd C$
in $G^*[D]$ since there is no $m$-connecting path between $i$ and $j$
given $C$
in $G^*$ and by removing nodes and edges from $G^*$ no new
$m$-connecting paths are generated. Therefore, by the induction
hypothesis $\langle i,j\cd C\rangle\in\alpha(\mathcal{J},M)$.
This implies that $\langle i,j\cd C\rangle\in\mathcal{J}$.

\emph{Case I.2}. Now suppose that $M=\varnothing$ and thus the node
set of $G^*$ is $D=\{i\}\cup\{j\}\cup\ant(i)\cup\ant(j)$. We prove
the result by reverse induction on $| C| $: For
the base, $C=V\setminus\{i,j\}=(\ant(i)\cup\ant(j))\setminus\{i,j\}
$ and the result follows trivially from the pairwise Markov property.

For the inductive step, consider a node $h\notin C$. We want to show
that $h$ is not simultaneously $m$-connected to both $i$ and $j$:
Suppose, for contradiction, there are $m$-connecting paths $\pi_1=\langle i,i_1,\dots,i_n,h\rangle$ and
$\pi_2=\langle h,j_m,j_{m-1},\dots,j_0=j\rangle$ given $C$. If (b1)
or (b2) of Lemma~\ref{lemj2} hold then $i$ and $j$ are $m$-connected
given $C$ which contradicts $i\perp_m j\cd C$. So we need only consider
the cases where $\langle i_n,h,j_m\rangle$ is collider or $i_n=j_m$
with an arrowhead
pointing to $h$ on the $i_nh$-edge. However, we know that $h\in\ant
(i)$ or $h\in\ant(j)$. Because of symmetry between $i$ and $j$
suppose that $h\in\ant(i)$.
%If $h\in\ant(i)\setminus\an(i)$ then by Lemmas \ref{lemvvn} and
%and $j$. If
Since $G^*$ is an anterior graph and there is an arrowhead pointing to
$h$ we have $h\in\an(i)$. Hence, there is a direction-preserving path
$\pi$ from $h$ to $i$. If no node on $\pi$ is in $C$ then (b1) or (b2)
of Lemma~\ref{lemj2} implies that the combination of $\pi$ and $\pi_2$ is an $m$-connecting path between $i$ and $j$, again a
contradiction. If there is a node on $\pi$ that is in $C$ then $h\in
\an(C)$ and again, by (a1) and (a2) of Lemma~\ref{lemj2}, $i$ and $j$
are $m$-connected given $C$, again a contradiction.

We conclude that, given $C$, $h$ is not $m$-connected to both $i$ and
$j$. By symmetry, suppose that $i\perp_m h\cd C$.

We also have that $i\perp_m j\cd C$. Since $\mathcal{J}_m(G^*)$ is a
compositional graphoid (Theorem~\ref{thm110}) the composition
property gives
that $i\perp_m \{j,h\}\cd C$. By weak union for $\perp_m$ we obtain
$i\perp_m j\cd \{h\}\cup C$ and $i\perp_m h\cd \{j\}\cup C$. By the
induction hypothesis,
we obtain $\langle i,j\cd \{h\}\cup C\rangle\in\mathcal{J}$ and
$\langle i,h\cd \{j\}\cup C\rangle\in\mathcal{J}$. By intersection,
we get
$\langle i,\{j,h\}\cd C\rangle\in\mathcal{J}$. By decomposition we
finally obtain $\langle i,j\cd C\rangle\in\mathcal{J}$.

\textit{Part II}. We now prove the result in the general case by
induction on $| C| $. The base, that is, the case that $| C| =0$, follows
from part I. To prove
the inductive step, we can assume that $C\nsubseteq\ant(i)\cup\ant
(j)$, since otherwise part I implies the result.

We first show that if $C\nsubseteq\ant(i)\cup\ant(j)$ then there is
a node $l$ in $C$ such that $i\perp_m j\cd C\setminus\{l\}$:
Let first $l'\in C\setminus(\ant(i)\cup\ant(j))$ be arbitrary. If
there is an $l''\in C\setminus(\ant(i)\cup\ant(j))$ so that $l'\in
\ant(l'')$ and $l''\notin\ant(l')$ then replace $l'$ by $l''$, and
repeat this process until it terminates, the latter being ensured by
transitivity of $\ant$ (Lemma~\ref{propve}) and the finiteness of
$C$. Thus, we eventually obtain an $l$ so that if $l\in\ant(\tilde
l)$ for $\tilde l\in C\setminus(\ant(i)\cup\ant(j))$ then we also
have $\tilde l\in\ant(l)$.

Suppose, for contradiction,
that there is a shortest $m$-connecting
path $\pi$ between $i$ and $j$ given $C\setminus\{l\}$. If $l$ is not
on $\pi$ or is a collider on $\pi$ then $\pi$ is also $m$-connecting
given $C$.
Therefore, $l$ is a non-collider on $\pi$. This, together with $l\notin\ant(i)\cup\ant(j)$, by using Lemma~\ref{lem11300}, implies
that $l$ is an
anterior of a collider node $p$ on $\pi$. Since $\pi$ is
$m$-connecting, $p\in C\cup\an(C)$. %If $p\notin\ant(l)$ then $p$ has
%to be in $\ant(i)\cup\ant(j)$,
%which implies that $l\in\ant(i)\cup\ant(j)$, a contradiction.
Thus, there is an\vspace*{2pt} $\tilde l\in C$ so that $p=\tilde l$ or $p\in\an
(\tilde l)$. Transitivity of anterior sets and the fact that $l\notin
(\ant(i)\cup\ant(j))$ now imply that $\tilde l\in C\setminus(\ant
(i)\cup\ant(j))$. The construction of $l$ implies $\tilde l\in\ant
(l)$ which again implies that $\tilde l\in\an(l)$ and $l\in\an
(\tilde l)$ and thus the collider tripath containing $p$ is a cyclic
ribbon unless
its endpoints are adjacent with an endpoint-identical edge, which
implies that $\pi$ is not a shortest $m$-connecting path, a contradiction.

We now have that either $i\perp_m l\cd C\setminus\{l\}$ or $j\perp_m
l\cd C\setminus\{l\}$
since otherwise, by Lemma~\ref{lemj2} there is an $m$-connecting path
between $i$ and $j$ given $C\setminus\{l\}$ in the case that $l$ is a
non-collider or given $C$ in the case that $l$ is a collider node.
Because of symmetry suppose that $i\perp_m l\cd C\setminus\{l\}$.
By the induction hypothesis, we have $\langle i,j\cd C\setminus\{l\}
\rangle\in\mathcal{J}$ and $\langle i,l\cd C\setminus\{l\}\rangle
\in\mathcal{J}$.
By the composition property we get $\langle i,\{j, l\}\cd C\setminus\{
l\}\rangle\in\mathcal{J}$. The weak union property implies
$\langle i,j\cd C\rangle\in\mathcal{J}$.
\end{pf}
If we specialise Theorem~\ref{thm113} to the most common
case of probabilistic independence models, we get the following corollary.
%
%co2 #&#
\begin{coro}
Let $G$ be a maximal ribbonless graph. A probabilistic independence
model that satisfies the intersection and composition axioms satisfies
the pairwise Markov property w.r.t. $G$ if and only if it satisfies the
global Markov property w.r.t. $G$.
\end{coro}
%
%s6.3 #&#
\subsection{Necessity of compositional graphoid axioms}
\begin{figure}[b]

\includegraphics{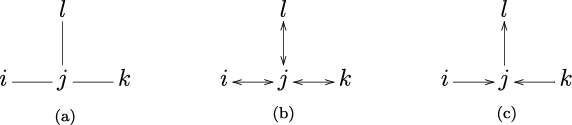}

\caption{For the equivalence of pairwise and global Markov
properties, (a) an undirected graph $G_1$ that shows that the
intersection property is necessary;
(b) a bidirected graph $G_2$ that shows that the composition property
is necessary; (c) a directed acyclic graph $G_3$ that shows that the
intersection property is necessary.}
\label{fig1114}
\end{figure}

Theorem~\ref
{thm113} states that, for equivalence of pairwise and global Markov
properties, the six compositional graphoid axioms are sufficient. In
fact, in general, for the mentioned equivalence, all six axioms are
also necessary. The graphs in Figure~\ref{fig1114} show that the
intersection and composition properties are necessary for the
equivalence of pairwise and global Markov properties.

For $G_1=(V_1,E_1)$, if $\mathcal{J}_1$ defined over $V_1$ satisfies
the pairwise Markov property, then $\langle i,k\cd \{j,l\}\rangle$,
$\langle i,l\cd \{j,k\}\rangle$, and $\langle k,l\cd \{i,j\}\rangle$
are in $\mathcal{J}_1$. It can be seen that none of the compositional
semi-graphoid axioms can be used to imply $\langle i,\{k,l\}\cd
j\rangle\in\mathcal{J}_1$. The intersection property is the only
axiom that implies the result.

For $G_2=(V_2,E_2)$, if $\mathcal{J}_2$ defined over $V_2$ satisfies
the pairwise Markov property then $\langle i,k\cd \varnothing\rangle
$, $\langle i,l\cd \varnothing\rangle$, and $\langle k,l\cd
\varnothing\rangle$ are in $\mathcal{J}_2$. It can be seen that none
of the graphoid axioms can be used to imply $\langle i,\{k,l\}\cd
\varnothing\rangle\in\mathcal{J}_2$. The composition property is
the only axiom that implies the result.

For $G_3=(V_3,E_3)$, if $\mathcal{J}_3$ defined over $V_3$ satisfies
the pairwise Markov property then $\langle i,k\cd \varnothing\rangle
$, $\langle i,l\cd \{j,k\}\rangle$, and $\langle k,l\cd \{i,j\}\rangle
$ are in $\mathcal{J}_3$. It can be seen that none of the
compositional semi-graphoid axioms can be used to imply $\langle l,\{
i,k\}\cd j\rangle\in\mathcal{J}_3$. The intersection property is the
only axiom that implies the result. See also for example, Example~3.26
of \cite{lau96}, showing that the pairwise Markov property does not
imply the global Markov property for DAGs when intersection is violated.
%
%f9 #&#

It is known that, for undirected graphs, the five graphoid axioms are
necessary and sufficient for equivalence of pairwise and global Markov
properties; see \cite{lau96}. For bidirected graphs, the independence
statement associated with a missing edge between nodes $i$ and $j$ is
$\langle i,j\cd \varnothing\rangle$ and only the five compositional
semi-graphoid axioms are necessary for equivalence of pairwise and
global Markov properties. This can be inferred from the proof of
Theorem~\ref{thm113}, since part I of the proof is not relevant for
bidirected graphs unless $C=\varnothing$ and the intersection property
is not used in part II of the proof. We conclude by stating this as its
own proposition.
%
%pr2 #&#
\begin{prop}\label{thmbidir}
Let $G=(V,E)$ be a bidirected graph. If an independence model $\mathcal
{J}$ defined over $V$ is a compositional semi-graphoid then $\mathcal
{J}$ satisfies the pairwise Markov property w.r.t. $G$ if and only if
it satisfies the global Markov property w.r.t. $G$.
\end{prop}

\section*{Acknowledgements}
We are grateful to Milan Studen\'y, Nanny Wermuth, and anonymous
referees for very helpful
comments on earlier versions of this paper.

%suskaldyti doi

% imsref loaded by jurgita.kaciuliene, 2013-02-06 10:23:11
% imsref loaded by vpetrauskas, 2013-12-30 09:48:42
% imsref loaded by vpetrauskas, 2013-12-30 09:56:44

\printhistory

\end{document}